\documentclass[acmsmall,screen,nonacm]{acmart}

\usepackage{booktabs}   %
\usepackage{subcaption} %

\input{setup}
\newcommand{\tfstep}[2]{\mathrel{\xrightarrow{#1}_{#2}}}

\begin{document}

\title{Formal Verification of an Out-of-Order Multiprocessor against an In-Order Weak-Memory ISA}

\author{Janggun Lee}
\orcid{0009-0002-0047-7717}
\affiliation{%
  \institution{KAIST}
  \city{Daejeon}
  \country{Republic of Korea}
}
\email{janggun.lee@kaist.ac.kr}

\author{Jeehoon Kang}
\orcid{0000-0002-2115-0871}
\affiliation{%
  \institution{FuriosaAI}
  \city{Seoul}
  \country{Republic of Korea}
}
\email{jeehoon.kang@furiosa.ai}

\begin{abstract}
  {\noindent
Out-of-order multiprocessors are critical components of modern hardware, and their verification is desirable as they are difficult to correct once shipped.
Existing literature simplifies and improves instruction set architectures (ISAs) for such processors, but they assume the processor correctly implements the ISA rather than proving it.

We present the first formal verification of an out-of-order multiprocessor \CoreL{} against an in-order, weak-memory \RISCV{} ISA, \RISCVPROM{}.
Our central component is \CoreIO{}, an in-order one-instruction at a time core specification, enabling us to confine out-of-order reasoning to a single core.
We decompose the main proof into two refinements.
The \emph{core refinement} proves that \CoreIO{} is a correct specification for \CoreL{}.
The \emph{system refinement} shows that \RISCVPROM{} is a correct specification for multiple \CoreIO{} connected with shared memory, leveraging the simplicity of \CoreIO{} in the proof.
All of our proofs are mechanized in Rocq, with large language model agents heavily used to automate proof writing.
We synthesized \CoreL{} into the Bluespec hardware description language, and validated it produces the intended weak behaviors and adheres to the official \RISCV{} test suite.
}
\end{abstract}

\begin{CCSXML}
  <ccs2012>
     <concept>
         <concept_id>10011007.10011006.10011008</concept_id>
         <concept_desc>Software and its engineering~General programming languages</concept_desc>
         <concept_significance>500</concept_significance>
         </concept>
   </ccs2012>
\end{CCSXML}

\ccsdesc[500]{Software and its engineering~General programming languages}

\maketitle

{\section{Introduction}
\label{sec:intro}

Multicore processors are fundamental components of modern systems, but they are difficult to design and reason about because they exhibit \emph{weak behaviors} that no sequentially consistent execution explains.
Such behaviors arise from the combination of two aspects of execution (see \cref{sec:bg-core} for concrete example).
First, the cores are subject to \emph{inter-core interleaving}, where the order in which their reads and writes may reach shared memory out-of-order.
Interleaving alone yields only \emph{sequential consistency} (SC), where every outcome is explained by some in-order execution of the cores.
Second, each core performs various optimizations, notably \emph{intra-core out-of-order execution}, in which independent instructions execute out of program order.
For example, a later load can execute before an earlier one whose address is not yet ready.

To capture such weak behaviors, modern specifications for instruction set architectures (ISAs) and programming languages employ a \emph{weak memory model}.
A rich literature simplifies and improves the design of such weak memory models~\cite{batty-C11,multicopy-arm,power,progress,promising-2,promising-arm}.

\parhead{Verification of processors}
Existing work on weak-memory semantics, however, assumes that a processor correctly implements its ISA rather than proving that it does.
It is very important that processors correctly implement their ISA because
\begin{enumerate*}
  \item they are complex, so a bug is likely; and
  \item a bug is costly, since a processor in production is difficult to update, unlike software
\end{enumerate*}.

Consequently, it is crucial to prove strong correctness theorems that gives us general results connecting real hardware and ISA.
Concretely, the theorems should be:
\begin{enumerate*}
  \item about synthesizable out-of-order multiprocessor designs;
  \item about established weak-memory ISA semantics; and
  \item parametric in hardware configuration, such as buffer sizes and core counts, and in execution length
\end{enumerate*}.
Existing results leave gaps in implementation coverage, ISA correspondence, or generality (see \cref{sec:related} for details).

Model checking of processor implementations targets bounded executions and components.
Some analyses bound execution length~\cite{arm-formal,arm-trustworthy} or check fixed set of litmus programs~\cite{pipecheck,rtl2uspec}.
\citet{jhala-mcmillan} verify unbounded executions of an out-of-order single-core model, but provide no synthesizable implementation.
QED combines reduction theorems with RTL model checking to verify ordering assertions in BOOM's load-store queue~\cite{qed} for arbitray programs, but is limited to a fixed queue configuration, and does not account for full system correctness.

Mechanized deductive verification of processors has focused on single-core or sequentially consistent models.
\citet{fm9801,completion-functions} cover speculative out-of-order single cores that does not perform load-load reordering, and \citet{kami,fjfj,revamp-verilog}, only consider in-order processors.
\citet{multiproc-deductive} mechanize a proof of an abstract speculative out-of-order multiprocessor, but it serializes stores, and their loads are revalidated in program order, which also serializes loads.

\citet{wmm} address weak-memory correctness for abstract out-of-order multiprocessors, but their proof argument is incomplete.
They propose in-order weak-memory specifications and give an informal argument that these models satisfy them.
An earlier arXiv preprint~\cite{i2e} presents simulation proofs for such models, but its stated invariants are insufficient to justify the required simulation steps (see \cref{app:i2e-proof-gaps}).

\subsection{Contributions}
\label{sec:intro:approach}

\textbf{We present the first formal verification of a synthesizable out-of-order multiprocessor against an in-order, weak-memory ISA.}
Formally, we prove the following theorem, where $n$ is the number of cores and $\sqsubseteq$ denotes simulation.
\begin{theorem}
\centering
\label{thm:main}
$\All n. n \cdot \CoreL{} + \DMem{} + \IMem{} \sqsubseteq \RISCVPROM{}$.
\end{theorem}
Our theorem connects a synthesizable core implementation \CoreL{}, composed with abstract data- and instruction-memory specifications \DMem{} and \IMem{}, to \RISCVPROM{}~\cite{promising-arm}.
The guarantee covers arbitrary programs and execution lengths, and is parametric over hardware configurations.

The main challenge is to explain the results of out-of-order execution through a substantially different in-order weak-memory ISA.
As mentioned above, out-of-order memory accesses produce non-sequential behavior,
but are not unconstrained and limited to those allowed by the ISA specifications.
For example, two loads to the same location must not observe a newer write followed by an older one in program order, even if the second load executed first.
The proof must show how the hardware's ordering checks and recovery leave only computations that can be reproduced together by a legal in-order ISA execution, while considering interleaving affect from other cores.

\parhead{Proof decomposition around \CoreIO{} (\cref{sec:overview})}
Our key idea is to use an \emph{in-order core specification}, \CoreIO{} (\cref{sec:overview:core-io}), to separate reasoning about out-of-order instruction execution into a \emph{single core} level proof.
We establish \cref{thm:main} by composing the \emph{system refinement} $n \cdot \CoreIO{} + \DMem{} + \IMem{} \sqsubseteq \RISCVPROM{}$ (\cref{sec:overview:system-proof})
with the
\emph{core refinement} $\CoreL{} \sqsubseteq \CoreIO{}$ (\cref{sec:overview:core-proof}).
\CoreIO{} executes instructions in-order, one instruction at a time.
The core refinement thus explains the results of out-of-order execution into an in-order one at a single-core level, independent of interference from other cores.
Utilizing \CoreIO{}, the system refinement solely focus on inter-core obligations of merging memory request/response into an internal transition, and relating store draining to the ISA's promises.

To simplify the core refinement, we introduce \CoreOoO{} (\cref{sec:overview:core-ooo}), an \emph{out-of-order core specification} that consolidates instruction information in a single program-order list.
It factors the proof into $\CoreL{} \sqsubseteq \CoreOoO{} \sqsubseteq \CoreIO{}$.
Positions in \CoreOoO{}'s list identify in-flight instructions, and flushing mis-speculation instruction can simply done by dropping a suffix.
This design simplifies the proof of $\CoreOoO{} \sqsubseteq \CoreIO{}$, which is a \emph{replay} argument constructing an in-order execution during instruciton commit.
\CoreOoO{} allows the proof to use list positions and suffix removal directly, without tracking distributed instruction information across various hardware modules.

\parhead{Modular proofs of various submodules (\cref{sec:proof})}
We prove the rest of the core refinement, $\CoreL{} \sqsubseteq \CoreOoO{}$, by further decomposing the proof by giving modular specifications for various submodules of $\CoreL{}$.
The main design choice is how to maintain speculation information, so that mispredicted branches/loads can easily be resolved across all modules in sync.
Each unresolved branch has a distinct identifier, and each instruction entry carries a \emph{speculation mask}, the set of identifiers of the branches on whose predictions it depends.
Recovery is then a simple filter operation that removes entries whose masks contain the mispredicted branch's identifier.

\parhead{Proof mechanization (\cref{sec:mechanization})}

We mechanize \cref{thm:main} in Rocq using Fjfj~\cite{fjfj}, a verification framework for a rule-based hardware description language inspired by Bluespec~\cite{bluespec}.
Large language model (LLM) agents serve as the proof workhorse.
The authors designed the initial specifications, invariants, and implementations, while the agents wrote the majority of proofs.

\parhead{Evaluation (\cref{sec:evaluation})}
Our implementation \CoreL{} is a heavily pipelined design inspired by RiscyOO~\cite{riscyOO}.
We translate our design into Bluespec for synthesis into RTL.
We ran our synthesized design against supported ISA tests in the \code{rv32ui} and \code{rv32ua} test bench~\cite{riscv_tests}%
\footnote{We do not support instruction fences and mixed sized accesses (half/quarter word loads and stores), whose behavior in promising semantics are unclear.
},
and additional litmus tests and confirmed the cores exhibit the desired weak behaviors.

\parhead{Limitations and future work}
Our verification covers \CoreL{} composed with the abstract memory specifications \DMem{} and \IMem{}, leaving cache implementations outside its scope.
We believe cache implementation and verification to be straightforward leveraging existing work~\cite{kami,hemiola,composable-coherence,cxl-coherence}.
\CoreL{} omits register renaming, scoreboarding, and superscalar execution.
We discuss directions for extending the design and verification to resolve these limitations in \cref{sec:related}.
}
{\section{Background}
\label{sec:background}

We review the out-of-order multiprocessor (\cref{sec:bg-core}), the promising semantics it implements (\cref{sec:bg-ps}), and our proof method, modular refinement (\cref{sec:bg-fjfj}).

We will use the program shown in \cref{fig:loadload} as a running example to guide our explanation.
The writer stores \code{1} to \code{x} and then \code{y}, both initially \code{0}, while the reader loads \code{y} into \code{a0} and then \code{x} into \code{a1}.
We will explain how the outcome $\code{a0}=1,\code{a1}=0$ is allowed when the addresses are distinct but forbidden when they are the same below.

\subsection{Design of the Out-of-Order Multiprocessor}
\label{sec:bg-core}

Our multiprocessor design connects multiple \CoreL{} with a data and instruction memory.
\CoreL{} implements the \RISCV{} base instruction set and atomic extension.
The cores are implemented in a rule-based hardware design language in the style of Bluespec~\cite{bluespec},
so a module consists of submodules, rules that update the submodule state, and external methods that interact with the outside world.
The design exhibits load--load, store--load, and store--store reordering.
The first two arise from the out-of-order load issue, and the last from the store buffer's out-of-order requests.

\Cref{fig:core-arch} shows the design of the core.
Blue rectangles indicate submodules, orange rounded boxes indicate internal rules, and purple rounded boxes with external arrows indicate the external methods.
We describe the core's methods, sketch how an instruction traverses the core, describe each component, and compose the cores into the full system.

\parhead{Methods}
A single core consists of five pair-wise request/response methods.
Instruction memory, plain loads and stores, and atomics, supports a request to shared memory and a response that arrives later.
External MMIO calls are lifted to requests for the top-level system's external behavior.
\begin{figure}
  \begin{minipage}{0.28\textwidth}
  \centering
  \begin{tabular}{@{}l||l@{}}
    \toprule
    \multicolumn{1}{c}{Writer} & \multicolumn{1}{c}{Reader} \\
    \midrule
    \code{x := 1} & \code{a0 = y} \\
    \code{y := 1} & \code{a1 = x} \\
    \bottomrule
  \end{tabular}
  \\
  \begin{tabular}{@{}lll@{}}
    \toprule
    \code{x} = \code{y}? & $\code{a0}=1,\code{a1}=0$ \\
    \midrule
    No & \greenCheck~allowed \\
    Yes    & \redX~forbidden \\
    \bottomrule
  \end{tabular}
  \caption{Reordering example with permitted outcomes. \code{x} and \code{y} are initially zero.}
  \label{fig:loadload}
  \end{minipage}
  \hfill
  \begin{minipage}{0.7\textwidth}
    \centering
    \includeinkscape[width=\textwidth]{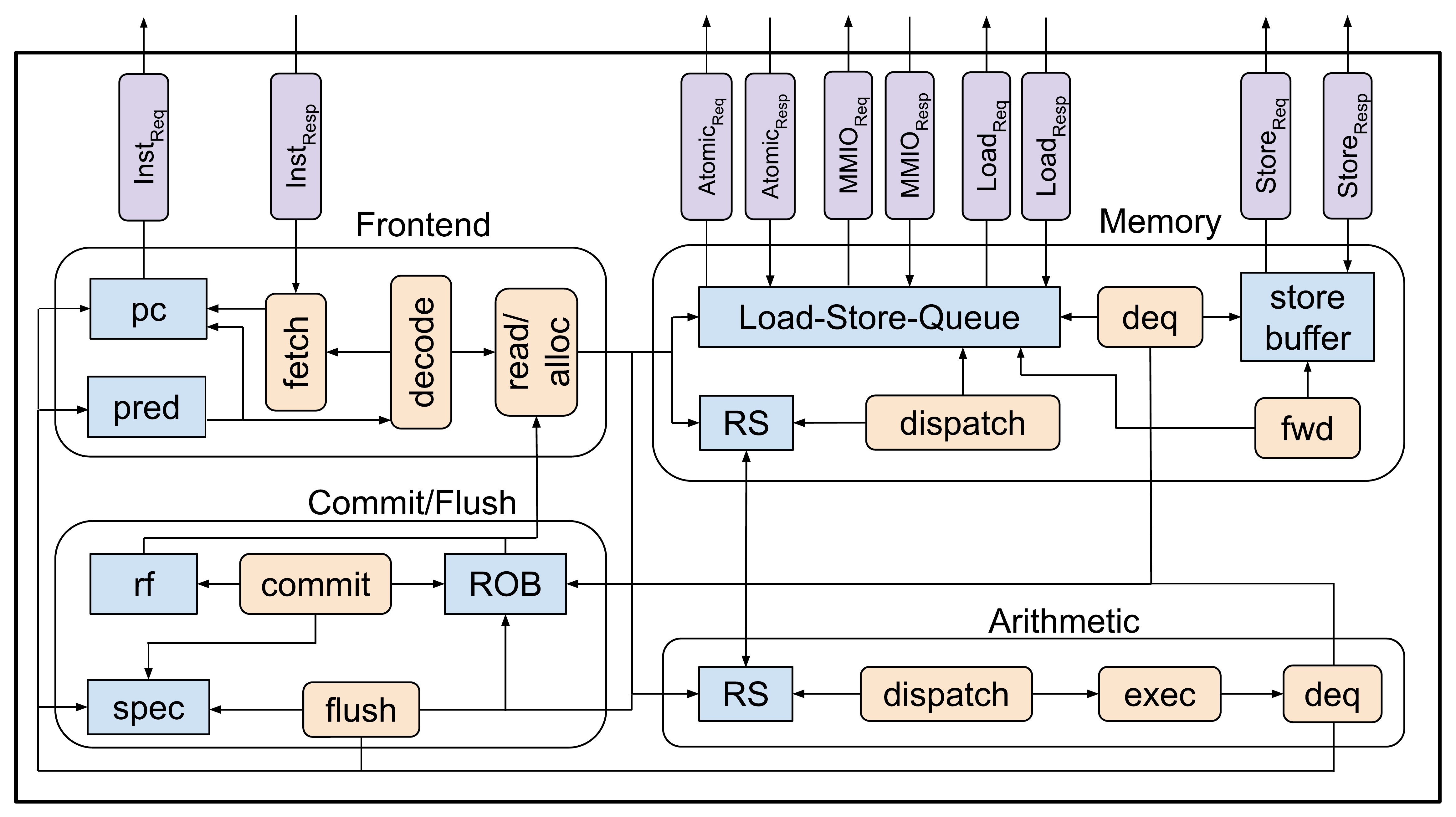}
    \caption{Design of the out-of-order core.
      Instructions flow from the PC \(\to\) Frontend \(\to\) Arithmetic/Memory \(\to\) Commit/Flush}
    \label{fig:core-arch}
  \end{minipage}
\end{figure}

\parhead{Instruction lifecycle}
Instructions are fetched and retired in program order, but may be dispatched out of order between these stages.
We follow the reader of \cref{fig:loadload} as it loads \code{y} and then \code{x}.
The core fetches each load from the instruction memory at the program counter, decodes it, and allocates an entry for it in the \emph{reorder buffer} (\Rob{}).
The core then reads the load's source operands, each represented as a value or a reference, and places the load in the memory unit's \emph{reservation station} (\Rs{}).
The reservation station holds the load until its operands are ready, then dispatches it for execution.
If the first load's address is not ready, the second load can be dispatched first.
Retirement from the \Rob{} proceeds in program order.

A computed load result, however, may need to be discarded before retirement.
In our running example, if the reader executes its second load before the writer's stores and its first load afterward, it obtains $\code{a0}=1,\code{a1}=0$.
In the same-address case, the core must flush the second load.

\parhead{Reorder buffer and register file}
The \Rob{} holds every in-flight instruction in program order, from allocation at decode to retirement, and serves as the core's dependency and speculation tracker.
The register file (\Rf{}) keeps the committed architectural state.
The \Rob{} resolves each source operand to one of three forms:
\begin{enumerate*}
\item a value from the register file, if no in-flight instruction writes to the register;
\item a value from the producing in-flight instruction, if it has finished executing in the \Rob{}; and
\item a reference to the producing in-flight instruction, if it is still executing in the \Rob{}.
\end{enumerate*}
The core tracks dependencies through \Rob{} indices, which identify each in-flight instruction.
The \Rob{} additionally tags every in-flight instruction with a \emph{speculation mask}, which is a bitmask that records which unresolved branches precede it.
Tracking speculation inside the \Rob{} is specific to our design (\cref{sec:proof:rob}), as it yields a cleaner specification.
Instructions retire from the \Rob{}'s head, writing their register results to the register file.

\parhead{Arithmetic unit}
The arithmetic logic unit (\Alu{}) executes \emph{pure} instructions, namely arithmetic instructions and branches, out of program order.
It selects any reservation-station entry whose operands are ready, computes the result, and broadcasts it to dependent instructions, all in a pipelined manner.
For a branch, the \Alu{} computes the actual direction, trains the branch predictor, and compares the direction against the prediction made during decode.
On a correct prediction, the core clears the branch from every younger instruction's speculation mask.
On a misprediction, the core flushes those instructions and redirects the fetch.

\parhead{Memory unit}
The memory unit (\Mem{}) issues loads speculatively and delays stores until after retire, using a program-order \emph{load--store queue} (\Lsq{}), a \Rs{}, and a \emph{store buffer} (\Stb{}).
As in \Alu{}, entries are marked as ready to issue requests out of program order.
A load either issues to memory or takes a forwarded value from an older in-flight store to the same address, either from the load--store queue or the store buffer.
Loads can issue if there are no older instructions to the same address, which is the source of load--load and store--load reorderings.
A store withholds its write until it retires, then moves into the \Stb{} and drains to memory.
\Stb{} drains stores to distinct addresses in no fixed order, making it the source of store--store reordering.
Atomic instructions are non-speculative and take effect only at the retire point.
An acquire fence blocks later loads from issuing, while a release fence stalls until \Stb{} is drained.
MMIO instructions are non-speculative, and loads cannot bypass it and stalls until \Stb{} is drained.

\parhead{Branch predictor}
The branch predictor supplies a predicted direction at fetch, letting the core speculate past an unresolved branch and keep fetching.
The \Alu{} trains the predictor whenever a branch's result is computed.

\parhead{Flushing excess loads}
The most important role of the \Lsq{} is to detect excess loads and mark them for flushing.
We illustrate the mechanism using the example of \cref{fig:loadload} in the same-address case.
Initially, the two loads are in the \Lsq{} in program order.
Suppose the second load to \code{x} has received its response, and the first load's address calculation has just finished.
We scan later instructions in the \Lsq{} and mark the second load for flushing because it may have read a value older than the first load's.
The load is flushed at a later time, after it is dequeued from the \Lsq{} and moves to the head of the \Rob{}.
When such a load is detected, the \Rob{} flags the core to flush the entire pipeline.

\parhead{Flushing mispredicted branches}
Mispredicted branches are eagerly flushed.
If a misspeculation is detected after execution, the core records the branch's speculation tag and flushes all instructions whose speculation masks contain that tag.

\parhead{Data and instruction memory}
The data and instruction memory each have matching interfaces to those of the core.
Instruction memory have interface for instruction request/response, and data memory have interface for loads, stores, and atomic requests.
They each additionally take a core identifier to distinguish requests from different cores.

\parhead{The multicore system}
The full system comprises multiple cores \CoreL{}, connected to $\DMem{}$ and $\IMem{}$.
Its request/response methods connect to the corresponding methods of each core,
and internal state consists of memory and bookkeeping for pending memory requests.
The system also exposes each core's MMIO as externally observable state.

\subsection{Promising Semantics}
\label{sec:bg-ps}

\begin{figure}
{\footnotesize \input{figure-promising-transitions}}
\caption{Types, internal transitions, and external methods of promising semantics.
We omit atomic operations for simplicity.}
\label{fig:promising-transitions}
\end{figure}

The promising semantics~\cite{promising,promising-arm} is an in-order, one instruction at a time operational semantics.
\cref{fig:promising-transitions} shows the types, internal transitions, and external methods.
The shared memory $H$ is not an array, but a history of all writes performed so far stored as a list.
Each core state ($cs$) consists of local state $\sigma$, set $P$ of \emph{promises}.
$\sigma$ consists of architectural states such as pc register, register file, and coherence information.
Coherence takes the form of a natural number representing ab index of $H$, and acts as a lower bound when loading from $H$.
A promise records the \emph{future stores} the corresponding core will make, and each element corresponds to the index of the store in $H$.
MMIO requests are stored in the finite partial map $IO$, with at most one request possible at a time per core.

The transition consists of single-core level transitions, and a lifting of the transition to full system execution.
\ruleref{Prom-Pure} executes pure operations (\eg, arithmetic, branch, jump, fence), thus only works on local state.
\ruleref{Prom-Load} executes a load instruction by selecting a store from the global shared memory $H$.
Here, a load reads from index $t$ of $H$.
The load transition $\sigma\tfstep{\msf{ld}(a,v), t}{}\sigma'$ checks that the load doesn't violate coherence conditions, via various lower bound conditions on $t$, and updates local state accordingly.
\ruleref{Prom-Promise} add a new write to the tail of $H$, and makes the core \emph{promise} that it will perform the write by adding the store's index to $P$.
As \ruleref{Prom-Promise} can add a write to global memory before the actual store executes, it can model load-store and store-store reordering.
\ruleref{Prom-Store} fulfills that promise, by executing a promised write at $t$.
\ruleref{Prom-Step} lifts the single-core transitions into the system-level transitions, and requires the new state to be \emph{certifiable}: the stepping core must be able to fulfill all its outstanding promises when executing in isolation.

For external methods, we have extended the semantics of \citet{promising-arm} with custom MMIO instrucitons.
\ruleref{ReqMMIO} call can be made when the current pc designates an MMIO instruction, there is no current request (\NONE), and the $\msf{empty }\, P$ predicate.
Here, $\msf{empty }\, P$ requires the promise set to be empty.
\ruleref{RespMMIO} call can be made when there is an active MMIO request ($\SOME\, q$).
\ruleref{Prom-MMIO} lifts the MMIO call to a full-system level call.

\subsection{Transitive and Compositional Refinement}
\label{sec:bg-fjfj}

We describe the notions of modules and refinement used throughout verification~\cite{fjfj}.
A module is a labeled transition system consisting of internal state, external interface, and internal transitions.
It is either primitive, with its transitions given directly as axiomatized specifications, or composed, built from submodules.

In Fjfj, an implementation module $M_i$ refines the specification module $M_s$, written $M_i \sqsubseteq M_s$, when there exists a simulation relation between their states such that
\begin{enumerate*}
  \item related states are indistinguishable through methods; and
  \item for every method of $M_i$, preserves the relation in lock step.
\end{enumerate*}
Refinement satisfies transitivity and compositionality.
\[
  M_1 \sqsubseteq M_2 \implies M_2 \sqsubseteq M_3 \implies M_1 \sqsubseteq M_3,
  \qquad
  M_i \sqsubseteq M_s \implies C[M_i] \sqsubseteq C[M_s],
\]
where $C$ is a module with submodule $M_i$.
Transitivity lets us build a refinement incrementally through intermediate specifications.
Compositionality permits freely replacing a submodule with its specification and vice versa.

Refinement imples traditional \emph{behavioral inclusion} ($M_i \subseteq M_s$), where behavior is a finite list of method calls.
Refinement implies inclusion between trace sets, which we write $M_i \subseteq M_s$.
Trace inclusion in Fjfj guarantees only safety, leaving liveness properties, such as eventual retirement, outside our scope.

}
{\section{The Core Specifications and Proofs to \RISCVPROM{}}
\label{sec:overview}

We present the in-order core specification \CoreIO{} (\cref{sec:overview:core-io});
its proof to \RISCVPROM{} (\cref{sec:overview:system-proof});
the out-of-order core specification \CoreOoO{} (\cref{sec:overview:core-ooo});
and the proof between the two core specifications (\cref{sec:overview:core-proof}).
The rest of the proof of $\CoreL{} \sqsubseteq \CoreOoO{}$ is presented in \cref{sec:proof}.

\subsection{In-Order Core Specification \CoreIO{}}
\label{sec:overview:core-io}

\begin{figure}
\raggedright
{\footnotesize \input{figure-core-io-transitions}}
\caption{Types, internal transitions, and methods for \CoreIO{}.
We omit state and methods for instruction memory and atomics for simplicity.}
\label{fig:core-io}
\end{figure}

\cref{fig:core-io} shows the types, transitions, and methods for the in-order core specification \CoreIO{}
Like a \RISCVPROM{} core (\cref{sec:bg-ps}), \CoreIO{} maintains a local state, which includes program counter, register file, and coherence information.
The store buffer $Q$, which records the executed stores in a per-address, program order list, replaces the promise set $P$.
The \emph{local memory history}, $LH$ records the core's loads responses and drained stores, and acts as a local subset of the global memory that this core has observed.
\CoreIO{} also carries storage for active load and store requests:
$L$ request the load identifier $LdIdx$ of all requests in a per-address list, where the list is stored in request order; and
$S$ records the store requests as a finite map from store identifier $StIdx$ to address.

The internal transitions utilize the same transitions as that of \RISCVPROM{}, with the main difference in loads and stores.
\ruleref{CoreInO-Pure} executes a pure instruction.
\ruleref{CoreInO-LoadFwd} forwards a load from the store buffer, and executes the load in the local state.
Here, the memory index is given as the length of local memory.
\ruleref{CoreInO-LoadMem} loads a memory from local memory, similarly to \ruleref{Prom-Load}.
The load from local memory can only be done when the corresponding store buffer is empty.
\ruleref{CoreInO-Store} executes a store, and adds the store to the core's store buffer.
\ruleref{CoreInO-Step} collects the transitions into a core-level transition.

The external methods are pairwise load and store requests, and those for MMIO.
\ruleref{CoreInO-ReqLd} requests a load for an arbitrary address $a$ at an arbitrary time.
Such freedom allows the core specification to match any speculative load requests from core implementations with that of specifications~\cite{fjfj}.
\ruleref{CoreInO-RespLd} serves the load response by adding the response value to the local memory $LH$.
\ruleref{CoreInO-ReqSt} requests a store for an address $a$ that has buffered stores.
\ruleref{CoreInO-RespSt} serves the store response by selecting an arbitrary store from the buffer, adding the store to the local memory, and ``draining'' all older stores for the same address.
\ruleref{CoreInO-MMIO} perfroms MMIO similar to \ruleref{Prom-MMIO}, re-using \ruleref{ReqMMIO} and \ruleref{RespMMIO}.
\CoreIO{}'s $\operatorname{empty}$ predicate requires $Q[a]$ as the empty list for all $a$.

\subsection{Proving $n \cdot \CoreIO{} + \DMem{} + \IMem{} \sqsubseteq \RISCVPROM{}$}
\label{sec:overview:system-proof}

We explain the specifications for \DMem{} and \IMem{}, for our intermediate top level module \TopMod{}, and specifications for
Decomposing the proof into two steps allows us to focus on the problems of atomizing memory interface and streamlining stores into promises separately.

\parhead{Specification for \DMem{} and \IMem{}}

\begin{figure}
  \centering
  {\footnotesize \input{figure-mem}}
  \begin{minipage}{0.49\textwidth}
  \centering
  \includeinkscape[width=0.26\textwidth]{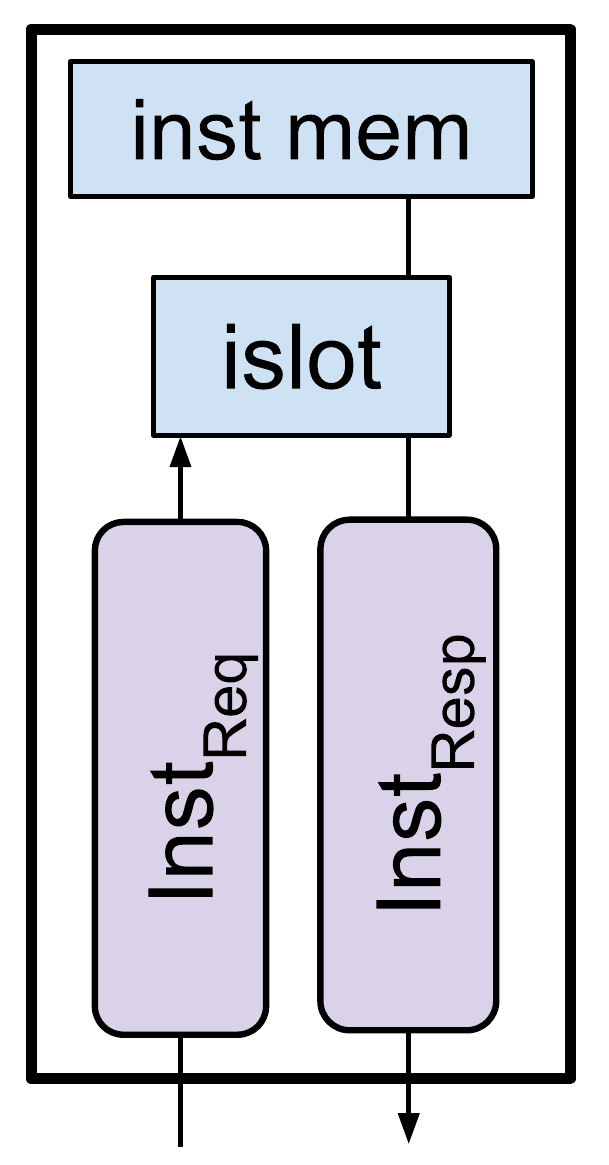}
  \quad
  \includeinkscape[width=0.37\textwidth]{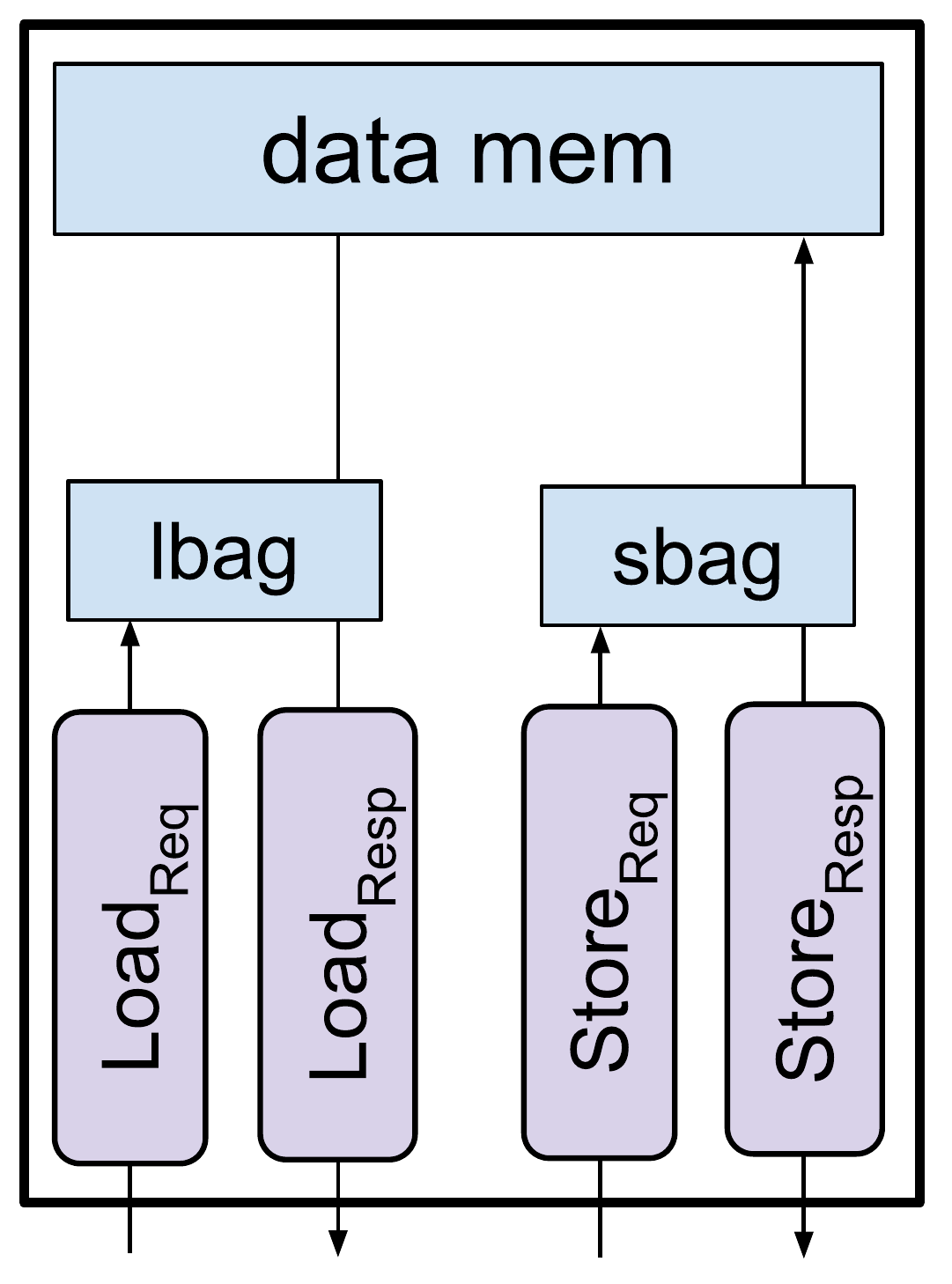}
  \caption{States and interface for inst and data memory.}
  \label{fig:mem-spec}
  \end{minipage}
\end{figure}

\cref{fig:mem-spec} describes the states and interface for the instruction and data memory.
The internal state is mostly a match of corresponding states to the \CoreIO{}, but indexed by the core identifier \TId{}.

The interface methods are standard: request methods takes a request with identifier for some core, request serves the request by consulting the instruction and data memory.

An important invariant for load response is that for each \TId{}, per-address requests are served in-order to ensure that intra-core coherence checks remain sound.
To see why this is needed, consider a simple programs with two loads to the same address, whose address calculation have both finished.
If the two load's request return out-of-order, the second load may have a value that is more stale than the first, trivially violating coherence.

\parhead{Specification for \TopMod{}}

\begin{figure}
{\footnotesize \input{figure-topmod-transitions}}
\caption{Types, internal transitions, and methods for the intermediate top-level module \TopMod{}.}
\label{fig:top-mod-spec}
\end{figure}

\TopMod{} (\cref{fig:top-mod-spec}) reuses \CoreIO{}'s entire buffered instruction relation, \mbox{$(cs,H) \tfstep{}{\msf{InO}} cs'$}.
Where \ruleref{CoreInO-Step} supplies a local history, \ruleref{TopMod-Step} supplies the shared write history and applies the relation to the selected core.
The same MMIO rules apply to that core and its slot.
A \CoreIO{} store response records only the last value of the prefix it drains at one address; \ruleref{TopMod-Drain} appends every value in that prefix to the shared history in order.

\cref{fig:top-mod-spec} describes the states, interface, and internal transition for \TopMod{}.
It simplifies memory interface transitions, and removes states related to them from \CoreIO{}, \DMem{}, and \IMem{}.
\ruleref{TopMod-Step} takes a step for a single core, re-using the $\xrightarrow{}_{\msf{InO}}$ transitions of \cref{fig:core-io}.
Note that here, $H$ is not the core's local memory, but shared global memory.
\ruleref{TopMod-Drain} selects a non-empty store buffer entry for a core, and drains it by appending it to the shared memory.
\ruleref{TopMod-MMIO} perfroms MMIO.
It is the same as \ruleref{Prom-MMIO}, with the sole difference the $\operatorname{empty}$ predicate, which is the same as that of \CoreIO{}.

\parhead{Proving $n \cdot \CoreIO{} + \DMem{} + \IMem{} \sqsubseteq \TopMod{}$}

Matching \CoreIO{}'s memory accesses in \TopMod{} requires translating coherence bounds from each core's local memory $LH$ to global memory $H$.
The simulation invariant carries a per-core $\msf{origin} : \LIST\,\mathbb{N}$ with length equal to the corresponding core's $LH$.
When $LH$ is extended with a newly observed value, \msf{origin} is extended with the index of the store supplying that value at the same address in \TopMod{}'s $H$ (with 0 for initial memory).
With this, the translation function is defined as $T_{cid}(k)=\max\bigl(\msf{take}(k,\msf{origin}_{cid})\bigr)$,
which takes the largest store index among the first $k$ origin indexes (with 0 for an empty list).
The prefix maximum makes $T_{cid}(k)$ monotonic in $k$.

The simulation invariant uses this translation to establish the coherence condition for a matching \TopMod{} load.
If $C_{\msf{InO}}$ and $C_{\msf{Top}}$ are corresponding \CoreIO{} and \TopMod{} cores, the invariant bounds the global coherence view by the translated local view:
$\All t,\msf{coh}. C_{\msf{Top}}.\msf{coh} \le T_{cid}\bigl(C_{\msf{InO}}.\msf{coh}\bigr)$ for any coherence state $.\msf{coh}$ of the cores.
If a load of \CoreIO{} reads address $a$ from a local entry at position $p$ with $C_{\msf{InO}}.\msf{coh}\le p$, monotonicity of $H$ gives
$C_{\msf{Top}}.\msf{coh} \le T_{cid}\bigl(C_{\msf{InO}}.\msf{coh}\bigr) \le H(p)$.
Thus \TopMod{} can read the same value at that origin while satisfying its coherence condition.

Other invariants preserve the correspondence of the remaining core state and memory interactions.
The store buffers contain corresponding values with related views, allowing forwarded loads to return the same value.
The relation also matches program counters and register values and tracks pending requests, MMIO and atomic operations, and reservations so that their transitions can be matched.

\parhead{Proving $\TopMod{} \sqsubseteq \RISCVPROM{}$}

The major challenge in this proof is that a \TopMod{} store transition cannot be matched lock-step by a \RISCVPROM{} store.
We keep the two systems' \code{data mem} equal so that loads can read corresponding writes.
Doing so requires store drains of \TopMod{} to be matched with promises of \RISCVPROM{}.
As \TopMod{}'s store executes \emph{before} it is drained from the buffer to \code{data mem}
whereas \RISCVPROM{}'s store executes (\ie, fulfilled) \emph{after} it is written to \code{data mem} with an earlier promise..
Thus, execution in \RISCVPROM{} lags behind \TopMod{}.

To bridge this timing gap, we maintain a per-core replay history inside the simulation invariant and use replay to make MMIO calls matchable.
During simulation, we record \TopMod{}'s internal steps in the history and match drained stores with promises.
MMIO calls must be matched immediately in \RISCVPROM{}.
Because \TopMod{} permits MMIO only with an empty store buffer, we replay whenever that buffer is empty, bringing \RISCVPROM{} up to date.

The replay invariant covers every possible drain order across cores and guarantees a hypothetical execution that catches up to \TopMod{}'s current core states.
For each order, the execution first promises the remaining buffered writes, then replays the recorded instructions without making further promises.
It ends with all promises fulfilled.
When a store drains, we make its promise in \RISCVPROM{} and retain all compatible replay executions with that promise already made.

The main remaining proof obligation is to certify the \RISCVPROM{} steps used to match drains and catch up.
Certification requires the core taking a step to be able to fulfill its remaining promises in isolation (\cref{sec:bg-ps}).
We adapt the backward-certification argument of \citet{promising-arm} to turn the hypothetical whole-machine completion supplied by replay into each core's independent certification, thereby justifying the actual matching steps.

\subsection{Out-of-Order Core Specification \CoreOoO{}}
\label{sec:overview:core-ooo}

\begin{figure}[t]
  \centering
  {\footnotesize \input{figure-core-ooo}}
  \includeinkscape[width=0.8\linewidth]{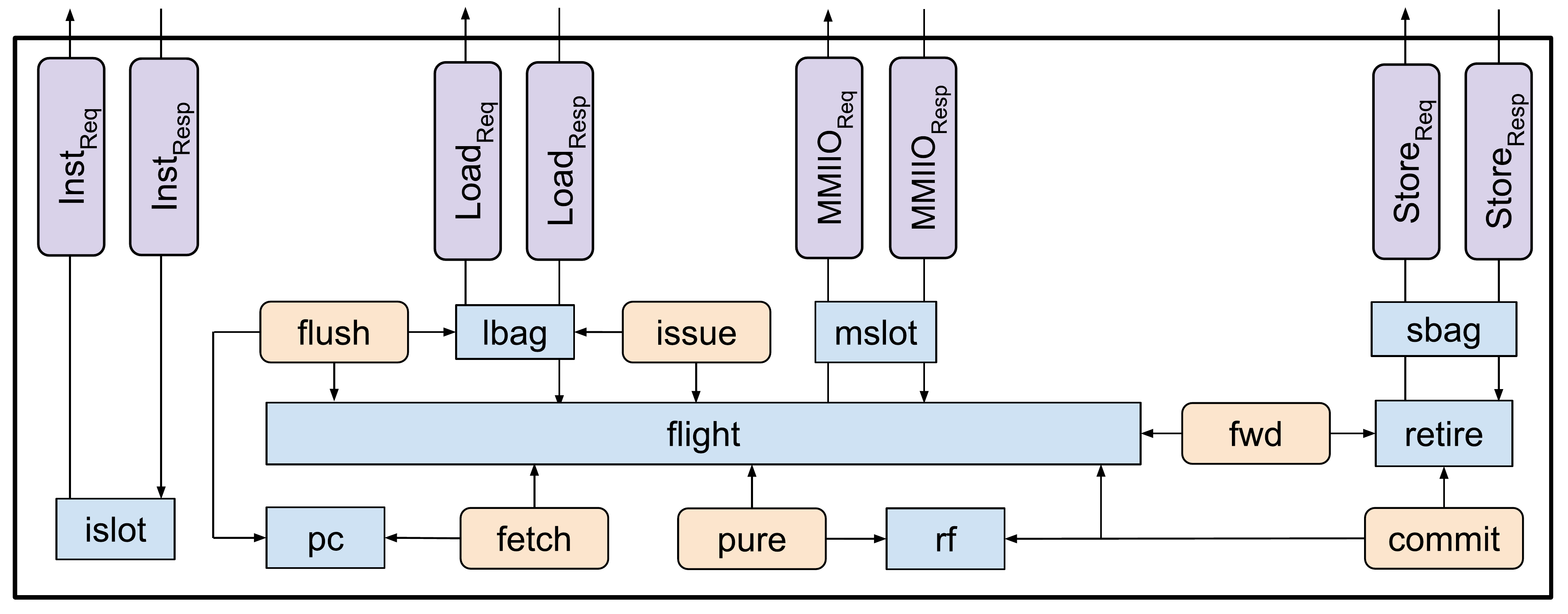}
  \label{fig:core-ooo:design}
  \caption{States, methods and transitions of \CoreOoO{}. We omit atomics for simplicity}
  \label{fig:core-ooo}
\end{figure}

\Cref{fig:core-ooo} describes how \CoreOoO{} specifies the core.
Compared to \CoreL{}, \CoreOoO{} abstracts away the mechanism for detecting misspeculation, the concrete causes of out-of-order execution, and duplicate sources of instruction metadata.
It leaves out-of-order execution, which is especially visible on how it deals with loads.

\parhead{State}
At the center is a program-order sequence of instructions consisting of a $\code{retire}$ prefix and a $\code{flight}$ suffix.
Retired instructions remain in $\code{retire}$ rather than being discarded.
Each in-flight and retired instruction therefore has the unique identifiers, $|\code{retire}| + \text{index in }\code{flight}$ and its index in $retire$, respectively.
We use these identifiers to label forwarding stores, which ables us to efficiently compute coherence violations.
Each $\code{retire}$ entry is either a pending store's write to memory (\code{WrMem}) or \NONE{}, and each $\code{flight}$ entry is an \msf{Inst}.
Other states are architectural state (\code{pc}, \code{rf}) and;
auxiliary state for memory transitions (\msf{lbag}, \msf{sbag}, \msf{mslot}, \msf{aslot}), similar to \CoreIO{} (\cref{sec:overview:core-io}).
The main difference lies in \msf{lbag}, which additionally tracks the identifier about in-flight loads in two queues, an issued but not yet requested queue ($\LIST\, \mathbb{N}$), and a request queue ($\LIST\, (LdIdx \times \msf{option}\, \mathbb{N})$).
A request entry whose instruction identifier is \NONE{} is not associated with a live load.

\parhead{Interface}
\CoreOoO{} has the same methods as \CoreIO{} and \CoreL{}.
Instruction, store, atomics, and MMIO requests proceed similar to that of \CoreIO{}.
We detail how \CoreOoO{} processes load request and response.
In a load request, we can unconditionally choose a random address, (for the same reason as in \CoreIO{} \cref{sec:overview:core-io}), or select one with a non-empty issue queue.
For either case, we allocate an id $lid$ for it and move it to the request queue, pushing $\NONE{}$ for the case of unconditional load.
In load response, we check if the originating load is still live.
If the load is in-flight, we records the response result into \code{flight}; and otherwise, simply discards the result.

\parhead{Internal transitions}
Instructions advance through the internal transitions shown in \cref{fig:core-ooo:design}.
We first \code{fetch} the current \code{pc}, add to the tail of \code{flight}, and update \code{pc}.
For branches and jump, we perform a random prediction for the next \code{pc}.
\code{pure} consists of the majority, updating instruction state in-place.
For instance, computation for an in-flight binary alu lookups the two source register values from earlier instructions with or \code{rf}, computes the results, and stores it in its instruction state.
We \code{commit} finished instructions from the \code{flight} head, updating \code{rf} and moving them to \code{retire}.
On a misspeculation, \code{flush} resolves it by dropping the tail of \code{flight} and updating \code{pc}.

Loads can also execute out-of-order.
A load with a ready address can be \code{fwd}ed from \code{retire}d stores or an earlier store in \code{flight}.
Otherwise, we \code{issue} the oldest ready, unissued load for that address by adding its identifier to \code{lbag}'s issue queue.
The rest proceeds by the interface methods.

\subsection{Proving $\CoreOoO{} \sqsubseteq \CoreIO{}$}
\label{sec:overview:core-proof}

The main challenge in this proof is matching ordinary (non-atomic/MMIO) instruction retirement in \CoreOoO{} with execution in \CoreIO{}.
As described in \cref{sec:overview:core-ooo}, retirement follows program order even when computation does not.
We match earlier computation steps while stalling \CoreIO{}, while matching memory requests and responses one-to-one.
Using \CoreIO{}'s nondeterministic load interface (\cref{sec:overview:core-io}), we record load responses in its local memory history $LH$.
The main invariants preserve register reads for in-flight instructions and establish load origins for completed loads.

The register read invariant ensures that when an ordinary instruction retires in \CoreOoO{}, the matching \CoreIO{} step reads the same register operands used in its earlier computation.
We require each computed operand to agree with the value supplied by the last preceding writer still in flight, or by the register file if no such writer remains.
At instruction retirement, there are no older in-flight instructions, so its operands must agree with \CoreIO{}'s registers.
If the retiring instruction writes a register, its value is transferred to the register file, maintaining the invariant for later instructions.

The load-origin invariant ensures that when a load retires in \CoreOoO{}, the matching \CoreIO{} step obtains the same value from its store buffer or local memory history.
We require each completed load's value to be justified by one of three origins: an older in-flight store, a buffered store, or an entry in local memory history.
Consider a load that forwards from an older in-flight store.
When the older store retires, it enters the buffer, moving the load's origin from the first to the second case.
When the store drains, its value is added to local memory history, moving to the third case.

For load origins in local memory history, the coherence invariants ensure that their positions satisfy \CoreIO{}'s coherence bounds.
For two completed loads $l_1$ and $l_2$ to the same address $a$, with local-history origins $o_1$ and $o_2$, we require $o_1 \le o_2$ whenever $l_1$ precedes $l_2$ in program order.
Thus, retiring $l_1$ does not raise \CoreIO{}'s coherence bound at $a$ above $o_2$, preserving the read for $l_2$.

}
{\section{Submodule Specifications and Proofs to \CoreOoO{}}
\label{sec:proof}

We describe specifications of comman modules (\cref{sec:proof:common}), and selective modules, namely
\Rob{} (\cref{sec:proof:rob}),
\Lsq{} (\cref{sec:proof:lsq}),
and
\Stb{} (\cref{sec:proof:store-buffer}),
and describe the remaining core refinement $\CoreL{} \sqsubseteq \CoreOoO{}$ (\cref{sec:proof:core-proof}).

Method specifications are given as triple $\hoare{P}{e}{Q}$ (see \cref{sec:precondition} for the exact definition).
In terms of using a method as a client, we must establish $P$ when calling $e$, and are guaranteed $Q$.
In terms of proving a method as a verifier, we can assume $P$, and can use it to establish $Q$.
In terms of Bluespec guards, as they are properties guaranteed by the callee, they are guarnteed in $Q$, not required by $P$.

\subsection{Common Building Blocks}
\label{sec:proof:common}

The development has three widely used libraries: sorted indexed list (\SIListT{}), the speculation-tag manager (\SMan{}), and reservation station (\Rs{}).

\begin{figure}
  {\footnotesize \input{figure-silist-spec}}
  \caption{Specification of the sorted indexed list \SIListT{}.}
  \label{fig:proof:spec-silist}
\end{figure}

\parhead{Sorted indexed list}
\Cref{fig:proof:spec-silist} specifies the sorted indexed list $\SIListT{}\ Idx\ T\ S$.
It is a list of elements in $T$, each carrying an ordered tag $S$, and the list is sorted by the tag.
Clients access elements through a stable index in $Idx$:
$idxs$ maps each stable index to an index in the list.
Each tag records the element's speculation state, \ie, older unresolved branches.
The well-formedness predicate $wf$ relates $list$ and $idxs$ and enforces the sort order.

The methods cover a typical queue lifecycle (\code{enq} and \code{deq}), a speculative kill, and a reset.
\ruleref{SIList-Enq} adds a new entry and returns a fresh index $idx$, given that the new tag keeps the list sorted (the $\msf{TLRel}$ premise).
\ruleref{SIList-Deq} removes and returns the first element of the list, along with its index.
\ruleref{SIList-KillIdx} applies $killIdx$, which finds the first element whose tag includes $sid$ and removes it and all later elements.
\ruleref{SIList-Clear} resets the list to a fresh state.

\begin{figure}
  {\footnotesize \input{figure-spectag-spec}}
  \caption{Specification of the speculation-tag manager \SMan{}.}
  \label{fig:proof:spec-spectag}
\end{figure}

\parhead{Speculation-tag manager}
\Cref{fig:proof:spec-spectag} specifies \SMan{} as a duplicate-free list of speculation tags with the newest at the head, where $SId$ is the tag type.
\ruleref{ST-Current} returns the current status of \SMan{} as a speculation mask, a bitset of $SId$s.
\ruleref{ST-Claim} allocates a new speculation tag and prepends it to the list.
\ruleref{ST-Correct} resolves a correct prediction by removing its tag $sid$ from the list.
\ruleref{ST-Wrong} either clears \SMan{} entirely ($k = \msf{All}$) or drops every tag from the head down to $sid$ ($k = \msf{One}\ sid$).

\begin{figure}
  {\footnotesize \input{figure-rs-spec}}
  \caption{Specification of the reservation station \Rs{}.}
  \label{fig:proof:spec-rs}
\end{figure}

\parhead{Reservation station}
\Cref{fig:proof:spec-rs} specifies \Rs{}, which holds each pending instruction's source operands until they are ready.
$\Rs{}\, Idx$ is a finite map from indices of type $Idx$ to reservation-station entries of type $\msf{RSE}$.
Each $\msf{RSE}$ holds two source operands ($r_1$, $r_2$) and the entry's speculation mask ($ss$), a set of values of type $SId$.
$\msf{RsSrc}$ either waits for the instruction with index $idx$ or holds a finished value $val$.

The methods insert and issue entries, snoop results into waiting operands, and resolve speculation.
\ruleref{Rs-Insert} adds a new entry to the reservation station.
\ruleref{Rs-Issue} removes and returns an entry whose source operands are both ready.
\ruleref{Rs-Snoop} finds the source operands waiting on $idx$ and marks them as ready with value $data$.
\ruleref{Rs-CorrectSpec} drops a correctly predicted branch's tag $sid$ from every entry's $ss$.
\ruleref{Rs-WrongSpec} clears the entire reservation station ($k = \msf{All}$), or removes entries dependent on a mispredicted branch ($k = \msf{One}\ sid$).

\subsection{The Reorder Buffer}
\label{sec:proof:rob}

\begin{figure}
  {\footnotesize \input{figure-rob-spec}}
  \caption{Specification of the reorder buffer \Rob{}.}
  \label{fig:proof:spec-rob}
\end{figure}

The \Rob{} allocates a slot for each fetched instruction, dequeues the instructions in program order, and in our design also manages speculation.
Whereas typical designs keep \SMan{} as a separate module, we embed it inside the \Rob{}, so the \Rob{}'s own methods allocate speculation tags.
The main benefit is that a client interacts with a single allocator module rather than coordinating several.
A client also need not maintain consistency between the internal states of \SMan{} and the buffer, because the \Rob{} refinement establishes that consistency.

\parhead{Specification}
\Cref{fig:proof:spec-rob} specifies \Rob{} as a \SIListT{}$\ Idx\ \msf{RobE}\ \msf{RobSS}$.
$\msf{RobE}$ records the destination register $dst$ and an execution state $ex$, \NONE{} until the instruction executes.
The execution state is either $Done$, carrying the register write (if any) and any program-counter redirect, or $IsStore$.
$\msf{RobSS}$ records two pieces of speculation information for an entry, its speculation mask ($ss$) and, for a branch entry, its own tag ($osid$).
The sort order precisely captures the speculation state.
For adjacent entries $e_1, e_2$, if $e_1$ is a branch with tag $sid$ then $e_2.ss = e_1.ss \cup \{sid\}$; otherwise $e_2.ss = e_1.ss$.

The methods combine speculation booking with queue operations.
\ruleref{Rob-Enq} adds a new instruction by allocating an index $idx$, and assigning a speculation mask $ss$ and, for a branch ($b = \code{true}$), a fresh tag $sid$.
It also ensures that if \Rob{} is empty, $ss$ is empty as well.
\ruleref{Rob-Deq} removes the head entry, given that an executed head has an empty $ss$, and guarantees that the dequeued entry has executed.
The empty-$ss$ condition holds because every branch before a committing instruction has already been resolved.
\ruleref{Rob-ReadReg} performs a ``register read'' on the \Rob{}, trying to obtain the data for the latest entry with destination register $reg$.

\parhead{Implementation}
The implementation composes $\SIListT{}\ Idx\ \msf{RobE}\ \msf{SSet}$ (the tag type is $\msf{SSet}$, not $\msf{RobSS}$) with \SMan{}, and each method forwards to the matching submodule methods.

\parhead{Proof}
The main invariant states that a tag lies in the implementation's \SMan{} exactly when it appears in the $osid$ of some entry in the specification \Rob{}.
The invariant also maintains monotonicity.
\ruleref{Rob-Enq} is the only hard case.
Its use of \ruleref{SIList-Enq} requires the \ruleref{ST-Current} postcondition together with the fact that every tag in \SMan{} already appears in the \Rob{}.
The other cases are routine.

\subsection{The Load--Store Queue}

\label{sec:proof:lsq}

\begin{figure}
  {\footnotesize \input{figure-lsq-spec}}
  \caption{Specification of the load--store queue \msf{LSQ} (state and commit methods).}
  \label{fig:proof:spec-lsq}
\end{figure}

\begin{figure}
  {\footnotesize \input{figure-lsq-spec-load}}
  \caption{Load methods of the load--store queue \msf{LSQ}.}
  \label{fig:proof:spec-lsq-load}
\end{figure}

The \Lsq{} is the core's most intricate module, with the most verbose interface and the most heavily pipelined internal state.

\parhead{Specification}
\Cref{fig:proof:spec-lsq} and \cref{fig:proof:spec-lsq-load} specify the \Lsq{}.
Its specification is similar to those of \CoreS{} and \Mem{}, but lacks a full $sbag$.
Its state consists of a \SIListT{} $in$ of in-flight memory operations, a load bag $lbag$ with an issued list and a requested list for each address, and a list $retire$ of retired writes.
Each entry, an $\msf{LSE}$, carries an at-commit flag and an operation.

We describe the methods of the \Lsq{}, omitting those shared with \SIListT{}.
We first follow the lifecycle of a load instruction (\cref{fig:proof:spec-lsq-load}), then describe the auxiliary methods (\cref{fig:proof:spec-lsq}).
\ruleref{LSQ-RdyLd} marks a load's address as computed.
It then kills every younger load to the same address that violates coherence, namely one that
\begin{enumerate*}
  \item has been issued,
  \item has received a response from memory, or
  \item has received a forwarded value from a store older than the just-readied load.
\end{enumerate*}
The store, load-reserved, and store-conditional resolutions are analogous and are omitted.
\ruleref{LSQ-IssueLd} issues a ready load when its value cannot be forwarded from the store buffer or an in-flight store in the LSQ, recording the request in $lbag$.
\ruleref{LSQ-IssueLd-Fwd} instead forwards from an in-flight store, completing the load in one step.
\ruleref{LSQ-ReqLd} carries the issued load to memory, moving its bag entry from the issued list to the requested list.
\ruleref{LSQ-RespLd-Live} brings the response back, updating the load to its response state.
The case of a response to a flushed load, which the queue discards, is omitted.
\ruleref{LSQ-DeqSt} retires a ready store into $retire$, but only if $ss$ is guaranteed to be empty.
\ruleref{LSQ-DeqRetire} pops the first committed write off $retire$.
\ruleref{LSQ-DeqNonSt} dequeues the committed head entry when it is not a store and $retire$ is empty.

\parhead{Implementation}
The implementation combines a \Rs{}, an $\SIListT{}\ LdIdx\ ((\msf{LSE}\ Idx) \times Idx)\ \msf{SSet}$.
\Rs{} dispatches ready operations, updates the corresponding \Lsq{} entries so that they can be requested
The implementation \Lsq{} is indexed by its own $LdIdx$ rather than the \Rob{}'s $Idx$, and each entry records the $Idx$ of its \Rob{} entry.

The implementation keeps no separate structure for the specification's $retire$ list.
A committed store instead stays in the \Lsq{} as a ready-to-retire entry, so $retire$ is a contiguous prefix of the \Lsq{} itself.
\ruleref{LSQ-DeqSt} marks the store ready in place rather than dequeuing it, and \ruleref{LSQ-DeqRetire} later pops the segment's oldest entry, which the memory unit passes on to the store buffer.

\parhead{Proof}
The difficult part of the proof is showing how \ruleref{LSQ-DeqSt} keeps a committed store in the implementation \Lsq{} while the specification dequeues it into $retire$.
The simulation relation has three main requirements:
\begin{enumerate*}
  \item the folded view of the reservation station and the implementation \Lsq{} equals the specification LSQ;
  \item the ready-to-retire prefix in the implementation LSQ, folded to its write effects, equals the specification's $retire$ and has an empty $ss$; and
  \item the map from $LdIdx$ to $Idx$ induced by the implementation \SIListT{} is injective
\end{enumerate*}.
The second invariant ensures that such ready-to-retire stores are not flushed by a mispredicted branch ($\msf{One}$).
When flushing a load, the implementation leaves these ready-to-retire entries intact.
These properties let us abstract the implementation's ready-to-retire stores and combine them with the \Stb{}.

\subsection{The Store Buffer}
\label{sec:proof:store-buffer}

\begin{figure}
  {\footnotesize \input{figure-stb-spec}}
  \caption{Specification of the store buffer SB.}
  \label{fig:proof:spec-stb}
\end{figure}

The store buffer (\Stb) collects retired stores and sends store requests to memory.
It responds to a request by coalescing the buffered writes to the requested address into a single memory write.

\parhead{Specification}
\Cref{fig:proof:spec-stb} specifies \Stb{} as a list $retire$ of committed writes and a finite map $sreq$ of requested addresses.
The length of $retire$ never decreases and its positions are never reused, so each position is a stable identifier that can record a load's forwarding origin.

The methods of the \Stb{} add new entries, send requests and responses, and check for forwarded values.
\ruleref{Stb-Fwd} scans $retire$ and returns the data for the last entry with a matching address, if any.
\ruleref{Stb-EnqSt} appends a store's write to $retire$.
\ruleref{Stb-EnqEmpty} instead appends an empty entry when a non-store instruction retires, keeping $retire$'s positions aligned with retirement order for the forwarding-origin bookkeeping.
\ruleref{Stb-ReqSt} make a request for a not yet requested address in $retire$.
\ruleref{Stb-RespSt} returns the address's last matching entry in $retire$, and replaces drained entries with \NONE{}.

\parhead{Implementation}
The implementation uses a single map module with key type $StIdx$, in which each element tracks an address, data, and issued status.
Only \code{enq\_st} is nontrivial, updating the entry that matches the store's address if one exists and allocating one otherwise.

\parhead{Proof}
The proof is relatively simple and relies on a relation with two conditions:
\begin{enumerate*}
  \item the specification's $sreq$ map is the implementation map with its non-issued entries dropped; and
  \item the value that $retire$ forwards for an address equals the data the implementation map stores for it
\end{enumerate*}.

\subsection{Proving $\CoreL{} \sqsubseteq \CoreOoO{}$}
\label{sec:proof:core-proof}

The main proof obligation is to show that when some internal transition of \CoreL{} takes place, \CoreOoO{} can take a matching step as well.
For most transitions, \CoreOoO{} is intentionally designed such that there is a matching transition, so the match can easily be made.
The simulation invariant mainly consists of trivial mappings of each \CoreL{}'s submodule's entry to the corresponding instruction in \CoreOoO{}:
each \Rob{} entry has a one-to-one correspondence to each in-flight instruction, and the microarchitectural states within execution units such (\ie, \Lsq{} and \Alu{}), and within matching instructions of \CoreOoO{} match as well.
We explain the non-trivial proof steps below.

\parhead{Preserving coherence information by tracking forwarding origins}
While both \Lsq{} and \CoreOoO{} simplify coherence computation by tracking instructions in a program order list, they are not identical as they store different information.
First, \Lsq{} tracks only memory instructions, while \CoreOoO{} tracks all instructions, so indexes of \Lsq{} are ``sparse''.
Second, a retired instruction in \Lsq{} may still be flushed, as a po-earlier branch may by flushed.
Thus, \Lsq{}'s retired instruction also contains information about flushed instructions.

We utilize two position maps to precisely map the forward index of \Lsq{} to that of \CoreOoO{} in a manner that preserves coherence information.
The first maps an index of in-flight \Lsq{} entry to an in-flight \CoreOoO{} entry.
The second maps an index of retired \Lsq{} entry to a retired or in-flight entry in \CoreOoO{}; and to \NONE if their entry has already been flushed.
With these maps, we map the forward index as follows.
If it is an in-flight store, we use the first map, and add the number of instructions the core has retired so far.
The main proof goal is showing that such maps stays consistent during instruction flush, which can add new entries to the second map mapping to \NONE.

\parhead{Matching fence completion}
A retired acquire fence in \Lsq{} causes a mismatch in load request availability in \CoreL{} and \CoreOoO{}.
When an acquire fence in retired from \Lsq{}, it no longer blocks younger loads in the \Lsq{} from being issued and requested.
However, as it is still inside \Rob{}, it is still in-flight inside \CoreOoO{}, so the same load cannot be issued nor requested in \CoreOoO{}.

We resolve this by introducing an intermediate core specification \CoreSF{}, where fences carry a ``finished'' boolean flag designating that they have dequeued from \Lsq{}.
In \CoreSF{}, an acquire fence with \code{true} finished flag doesn't block younger loads.
Thus, retirement of fences from \Lsq{} can be matched by a \CoreSF{} transition where the finished flag flips from \code{false} to \code{true}.

In the proof of $\CoreSF{} \sqsubseteq \CoreOoO{}$, the mismatch in fence semantics is resolved by a simulation invariant where we eagerly retire \CoreOoO{} as much as possible.
A property of fences is they are retired from \Lsq{} only when they are the first instruction of \Rob{}, hence first instruction of \CoreSF{}.
Thus, when a fence is marked as finished in \CoreSF{}, we can simply commit the instruction in \CoreOoO{}, since there are no earlier in-flight instructions blocking commit.

}
{\section{Proof Mechanization}
\label{sec:mechanization}
We report on the mechanization of the end-to-end refinement (\cref{sec:eval:mech}) and our experience proving it with LLM agents (\cref{sec:eval:llm}).
\subsection{Rocq Development}
\label{sec:eval:mech}
We mechanized the end-to-end refinement of \cref{thm:main} in Rocq using the Fjfj verification framework~\cite{fjfj}.
The theorem is parametric in the number of cores, buffer and map sizes, value bitwidth, and instruction decoder.

\subsection{LLM Agents as the Proof Workhorse}
\label{sec:eval:llm}
LLM agents wrote most of the proof scripts.
We used Claude and Codex agents, with Fable and Astra for planning and coordination and Sonnet and Luna for actual implementation.

\parhead{Workflow}
We worked with agents on all tasks, including implementing hardware modules, actual refinement proofs, and refactorings.
For each task, we sketched the intended result and refined the plan with an agent before asking it to carry out the work.
Changes to the agreed hardware design or specifications during implementaion required our approval.
Proof invariants or helper lemma statements was allowed to be freely changed, as they do not affect the end statement being proved.

As an example, we describe how we extended \citet{promising-arm}'s semantics with a program counter and instruction decoder.
We requested to remove ARM-specific features, explained the high-level design to add a program counter and instruction decoding, and gave the transition for one instruction as an example.
We also requested to record the original source alongside each refactored definition.
After the agent implemented the task, we manually reviewed roughly 500 lines of generated Rocq definitions with the original development.

\parhead{Specification bug of \CoreOoO{} found by agents}
An agent found that our initial treatment of loads in \CoreOoO{} lost information needed for the coherence.
Originally, we had ommitted away \msf{Issue}, an state for loads denoting it is to-be requested, to decrease the number of intermediate states.
Agents noted this made coherence calculation imprecise, showed examples where \CoreOoO{} exposed behaviors not allowed by \CoreIO{},
and correctly proposed exposing \msf{Issue} as the fix.

\parhead{Necessity of human guidance: adding atomics}
When extending \RISCVPROM{} with the atomic read-modify-write instructions of Zaamo, the agent initially appended each atomic write to shared memory when the instruction executed.
In the in-order ISA, this placed the write after the core's earlier stores, whereas the implementation can make a relaxed atomic write visible while an earlier store to another address is still buffered.
We suggested following the treatment of promises for successful store-conditional instructions.
The agent adapted that rule so the core could promise the atomic write in advance and fulfill it when the instruction executed.

}
{\section{Evaluation}
\label{sec:evaluation}
To synthesize that our design, we developed a simple Bluespec-like language that generates matching Bluespec~\cite{bluespec} and Fjfj implementations, and implemented \CoreL{} inside this new language, leading to a synthesizable implementation of \CoreL{} with a matching verifiable implementation.

\parhead{Translating to Bluespec}
Our language closely matches Bluespec in syntax, with additional information needed for Fjfj.
This significantly simplifies generation to Bluespec, as direct translation from Fjfj to Bluespec is non-trivial as Fjfj does not differentiate guards and method bodies.
Translation to Bluespec is straightforward, and to Fjfj is simple as well as we simply append the guard body and method body.
Additionally, having a source language similar to Bluespec allows us to implement \CoreL{} and other submodules in a Bluespec style-manner.

\parhead{Testing synthesized \CoreL{}}
The generated core passed the tests for supported instructions in the official \RISCV{} test suite~\cite{riscv_tests} and exhibites the desired weak behaviors
Specifically, it passes \code{rv32ui} tests for the base integer instruction set, (excluding those with instruction fences or subword memory accesses) and the \code{rv32ua} tests for the atomic extension.
In two-core litmus tests, we observed weak outcomes for load--load, store--load, and store--store reordering.
Each test used a memory schedule chosen to expose the intended reordering and, and is observed via MMIO.

}
{\section{Related and Future Work}
\label{sec:related}

We expand on the discussion of related work and limitations done in \cref{sec:intro}.

\parhead{Model checking processors}
Model checking has established properties of processor control and memory ordering for both abstract models and concrete RTL, but prove only bounded properties or does not achieve full system verification.
\citet{burch-dill} verified processor control by flushing the pipeline into an architectural state.
Subsequent work reached realistic designs through compositional model checking of Tomasulo's algorithm and richer microarchitectures~\cite{mcmillan-tomasulo,jhala-mcmillan}.
RTL2$\mu$SPEC~\cite{rtl2uspec} derives microarchitectural ordering specifications from RTL and was evaluated on a four-core system with three-stage in-order cores under SC.

\parhead{Formally verified processors}
Existing processor verification is limited to in-order processors or SC semantics.
Fox~\cite{fox-arm6} verified the ARM6 processor in HOL, covering a single in-order core with a three-stage pipeline.
\citet{fm9801,completion-functions} verified an out-of-order design with speculative execution, exceptions, and self-modifying code, but the design is limited to a single core.
\citet{multiproc-deductive} mechanized the verification of an out-of-order multiprocessor against an in-order one in Rocq.
However, the verified semantics executes loads in order, yielding SC semantics.
K\^{o}ika~\cite{koika} is a DSL that models Bluespec's scheduling semantics.
Its authors prove that this scheduling semantics is one-rule-at-a-time, and verify a single in-order core.
PipeProof~\cite{pipeproof} performs unbounded verification formulated as a SAT instance, often requiring manually supplied invariants.
However, they only consider in-order processors.

\parhead{Weak-memory semantics and its clients}
Our work extends the lower bound existing verification stack for weak-memory systems from ISA semantics to RTL designs for realistic, efficient multiprocessors.
For hardware memory models, \citet{herding-cats} comprehensively tested processor behavior against axiomatic models, including that of the official ARM specification.
\citet{multicopy-arm} presents the Flat memory model, which models ARMv8's in an abstract microarchitectural style similar to \CoreOoO{}, simplifying the interaction between individual cores and shared memory.
Our core specification \CoreOoO{} uses a single program-order list, whereas the Flat model represents instructions as trees.
The list lets us detect coherence violations locally by comparing instruction indices, while Flat requires global calculation.
\citet{promising-arm,i2e} both present an operational, one-instruction at a time model for weak semantics.
Compared to \CoreIO{}, which give specifications to a core independent of shared memory, their semantics is monolithic, in particular for those related to load and stores.
\citet{i2e}'s semantics is furthermore not core-local, as the draining of stores from one core affects that state of all other cores.
Compiler correctness proofs~\cite{promising-seq,promising-ir} connect hardware memory models to software weak-memory models.
Program logics~\cite{compass, weak-proof-recipie} and library verification~\cite{weak-smr,weak-reachability} extend the stack to user-level libraries.

\parhead{Limitations and future work}
We plan to implement a coherence cache system and proving refinement against \DMem{} and \IMem{}, and believe it to be straightforward leveraging existing work.
By utilizing compositional refinement (\cref{sec:bg-fjfj}), such improvement can be done in a black-box manner without touching existing proofs.
Since prior work~\cite{kami,hemiola,composable-coherence,cxl-coherence} already addresses much of cache-coherence verification, leveraging to our implementation should be straightforward.

We plan to improve to a more optimized design, namely implementing superscalar features such as register renaming and scoreboarding.
The main goal of superscalar executions is executing multiple instructions in one step.
Thus, one step of superscalar executions can be matched by multiple steps of our current core design, so we expect the main invariants and proof obligations to not change.
Notably, \CoreOoO{} should still remain a valid specification for a superscalar design, thus confining the proof updates within $\CoreL{} \sqsubseteq \CoreOoO{}$.

We follow design of modern cores~\cite{riscyOO, boom, xiangshan, promising-ir}, which omits load-store reordering as they barely provide a hardware optimization chance and complicates recovery.
Our target ISA of \RISCVPROM{} supports load-store and store-store reordering uniformly with promises, and matching a \TopMod{} store-store reordering with promises was one of the major proof difficulties (\cref{sec:overview:system-proof}).
There are ongoing discussions about revising ISA semantics to eliminate load-store reordering and simplify the design; it will be interesting to see how such design change can simplify our proofs.

}

\bibliography{reference}

\clearpage
\appendix
\section{Gaps in the I2E Simulation Arguments}
\label{app:i2e-proof-gaps}

\citet{wmm,i2e} describes the Instantaneous Instruction Execution (I2E) framework,
which describes weak-memory behavior using processors that execute instructions instantaneously in program order.
WMM's shared memory is modeled as a simple array, and has per-core store and invalidation buffers to model out-of-order memory accesses.
Store buffers buffer stores between processors and shared memory, modeling store-store reordering.
When a store is drained from the store buffer to shared memory, it is placed inside the invalidation buffer of every other core.
A load from a core can read from
\begin{enumerate*}
  \item the core's store buffer;
  \item the core's invalidation buffer; if the store buffer doesn't match; and
  \item shared memory, if the invalidation buffer doesn't match
\end{enumerate*}.
To preserve load-load coherence, when a load is read from the invalidation buffer, all older same-address stores inside the buffer are ``invalidated''.

Theorem 1 of \citet{i2e} argues that $\mathrm{CCM}+\mathrm{OOO}^{\mathrm{VP}}$, the composition a cache coherent module system and ab out-of-order core with load-value prediction, is behaviorally included in WMM.
(note that they do not formally define the meaning of behavioral inclusion nor modules, so we assume definitions that are similar to that of \citet{fjfj}).
The argument is done by setting up a simulation invariant, with the main components as follows:
\begin{enumerate*}
  \item the shard memory and store buffers of $\mathrm{CCM}+\mathrm{OOO}^{\mathrm{VP}}$ and WMM are equal; and
  \item every instruction committed in $\mathrm{CCM}+\mathrm{OOO}^{\mathrm{VP}}$ is executed in WMM with the same result
\end{enumerate*}.
Under these invariants, the main proof goal is to show that when we commit an instruction in hardware, we can commit in WMM, and when we drain a store in hardware, we can similarly drain it in WMM.
The main case is a load whose supplying write has already been overwritten in shared memory, and the proof argues that the overwrite inserts its value into WMM's invalidation buffer and that intervening instructions cannot remove it, using hardware coherence to handle intervening loads.

\begin{figure}
  \centering
  \small
  \begin{tabular}{@{}lll@{}}
    \toprule
    Reader & Publisher & Interferer \\
    \midrule
    $A:\ r1 := \text{ld } y$ & $S_1:\ \text{st }\ x\ 1$ & $S_3:\ \text{st }\ x\ 3$ \\
    $B:\ r2 := \text{ld } x$ & $S_2:\ \text{st }\ x\ 2$ & \\
    $C:\ r3 := \text{ld } x$ & $T :\ \text{ st }\ y\ 1$ & \\
    \bottomrule
  \end{tabular}
  \medskip

  \begin{tabular}{@{}lll@{}}
    \toprule
    & Hardware & WMM \\
    \midrule
    Memory-write order & $S_1,S_3,S_2,T$ & $S_3,S_1,S_2,T$ \\
    Reader execution order & $B=1,C=3,A=1$ & $A=1,B=1$ \\
    \bottomrule
  \end{tabular}
  \caption{Hardware and WMM states that satisfy Theorem 1's invaraints, but cannot proceed in simulation.}
  \label{fig:i2e-history}
\end{figure}

However, the main invariants of Theorem 1 are too weak to perform simulation proofs and allows intermediate states where the proof cannot proceed, since the invariants do not relate the performed stores of the hardware and WMM,
\cref{fig:i2e-history} illustrates sucah a program and proof states, namely a state pair where the hardware can commit an instruction but WMM cannot execute with the same result.
Specifically, the invariant allows the scenario where the store orders of $\mathrm{CCM}+\mathrm{OOO}^{\mathrm{VP}}$ ($S_1,S_3,S_2,T$) and WMM ($S_3,S_1,S_2,T$) differ, while having equal shared memory and store buffers (as all empty),
and the reader has committed $A$, and has executed $B$ and $C$ but is yet to commit them.
The different memory-write orders place the store of $3$ by $S_3$ before $S_1$ in the reader's WMM's invalidation buffer.
When we commit $B$ in hardware, the only way to match in $WMM$ is by executing $B$ with result $1$ from $S_1$ in WMM, which invalidates $S_3$.
Hardware can now commit $C$ with result $3$, but WMM can obtain only $1$ from its invalidation buffer or $2$ from memory, hence cannot produce $3$.
\citet{i2e}'s proof incorrectly restricts the store orders of hardware and WMM during so that such an example doesn't arise.
This is not to say that the statement of Theorem 1 is false, but that the stated invariants are too weak to fully establish the correctness proof.

\section{Modular Refinement with Preconditions}
\label{sec:precondition}

We generalize the undefined-behavior flag used in the Fjfj switch example~\cite{fjfj-artifact} by lifting specification states to \code{option}.
The uniform form of the lifted specifications and simulation relations supports reusable proof tactics.

Our key idea is to lift the specification state to \code{option}.
We review the per-method proof obligation (\cref{sec:precondition:background}), show why it admits no assumptions about arguments (\cref{sec:precondition:problem}), and present our encoding and its comparison with CCR (\cref{sec:precondition:method}).

\subsection{Background: Per-Method Proof Obligations}
\label{sec:precondition:background}
Fjfj reduces a refinement proof (\cref{sec:bg-fjfj}) to one proof obligation per method and rule.
Rules need no precondition because they have no caller who could establish one, whereas methods do.
We therefore focus on the proof obligations for methods.
A method's specification is a relation $spec(arg, st, st', ret)$ over the argument, the pre- and post-states, and the return value.
For a simulation relation $\phi$, the proof obligation for a method is
\[
  \begin{aligned}
    & \All arg, ret, st_i, st_i', st_s.\, \phi(st_i, st_s) \to spec_i(arg, st_i, st_i', ret) \to \\
    & \Exists st_s'.\ spec_s(arg, st_s, st_s', ret) \wedge \phi(st_i', st_s')
  \end{aligned}.
\]
We must show that if the initial states are related by $\phi$ and the implementation can take a step satisfying $spec_i$, then the specification can take a corresponding step satisfying $spec_s$ that preserves $\phi$.

\subsection{Problem: Unconstrained Arguments}
\label{sec:precondition:problem}

The proof obligation leaves no way to assume anything about a method's arguments.
Consider a module whose state is a finite map $st$, with a method $\code{update}(idx, f)$ that applies $f$ to the value at $idx$ and returns unit.
An ideal specification for \code{update} would be
\[
  spec_{\code{update}}((idx, f), st, st',\_) := \Exists v.\ st[idx] = v \wedge st' = st[idx \mapsto f(v)].
\]
However, the simulation proof requires this specification even for an $idx$ outside the map's domain, where no such $v$ exists.

The simulation proof needs the missing precondition as a hypothesis in its proof obligation.
\[
\begin{aligned}
  & \phi(st_i, st_s) \to spec_i((idx,f), st_i, st_i') \to (\Exists v.\ st_s[idx] = v) \to \\
  & \Exists st_s'.\ spec_{\code{update}}((idx,f), st_s, st_s',\_) \wedge \phi(st_i', st_s')
\end{aligned}
\]
A caller of \code{update} must then prove that the condition holds at the call site.

\subsection{Solution: Lifting State to \code{option}}
\label{sec:precondition:method}

Our lightweight encoding provides a limited form of the precondition mechanism of CCR and its successors~\cite{ccr,ccr-2,cris}, leaving the underlying semantics untouched.
CCR realizes conditions as wrappers, where a violated precondition triggers undefined behavior in the semantics.
Our encoding instead makes the specification state vacuous by setting it to \NONE{}.
CCR conditions own separation-logic resources and transfer them between caller and callee, while our preconditions are limited to pure predicates.
This restriction makes the simple encoding sufficient.

\parhead{Guarded specifications}
Our encoding separates a method specification into a precondition and a postcondition.
A \emph{guarded specification} pairs them over an argument type $A$, a state type $ST$, and a return type $R$.
\[
  \msf{GSpec}\ A\ ST\ R := \{\; \mathit{pre} : A \to ST \to \msf{Prop};\quad \mathit{post} : A \to ST \to ST \to R \to \msf{Prop} \;\}.
\]
The guarded specification for \code{update} is
\[
  \mathit{gspec}_{\code{update}} := \{\; \mathit{pre}((idx, f), st) := \Exists v.\ st[idx] = v;\quad \mathit{post} := spec_{\code{update}} \;\}.
\]
The figures in \cref{sec:proof} render a guarded specification as a Hoare triple, with $\mathit{pre}$ as the premise and $\mathit{post}$ as the postcondition.

\parhead{Lifting to \code{option}}
We lift both the guarded specification and the simulation relation to \code{option}, so a method call violating the precondition can simply update the internal state to \NONE{}.
A \msf{GSpec} becomes a method specification over $\code{option}\ ST$ as follows:
\[
  \begin{aligned}
    \msf{lift}(gspec)(arg, ost, ost', ret) :=
    & \All st.\ ost = \SOME\ st \to \mathit{pre}(arg, st) \to \\
    & \Exists st'.\ ost' = \SOME\ st' \wedge \mathit{post}(arg, st, st', ret) \\
  \end{aligned}
\]
The simulation relation $\phi$ is lifted similarly.
\[
  \msf{lift}(\phi) := \lambda\, st_i\ ost_s.\; \All st_s.\ ost_s = \SOME\ st_s \to \phi(st_i, st_s).
\]
Both lifted forms hold vacuously at \NONE.
A value method's specification is lifted in the same way, with a postcondition relating only the argument, the state, and the return value.

\parhead{Obtaining the precondition}
The guarded specification lets us use the precondition in the proof.
The proof obligation for \code{update} is now expressed over the lifted state:
\[
  \begin{aligned}
    & \All (idx,f), st_i, st_i', ost_s.\
      impl((idx,f), st_i, st_i') \to \msf{lift}(\phi)(st_i, ost_s) \to \\
    & \Exists ost_s'.\ \msf{lift}(\mathit{gspec}_{\code{update}})((idx,f), ost_s, ost_s') \wedge \msf{lift}(\phi)(st_i', ost_s')
  \end{aligned}
\]
We perform case analysis on $ost_s$ and then on the validity of $pre$.
If $ost_s = \NONE$, choosing $ost_s' = \NONE$ trivially solves the obligation.
If $ost_s = \SOME\ st_s$ but $pre((idx,f),st_s)$ does not hold, the proof again chooses $ost_s' = \NONE$.
To prove $lift(gspec_{\code{update}})$ in this case, we assume its premise $pre((idx,f),st_s)$, yielding a contradiction.
The remaining case is when the precondition holds, and the goal simplifies to the ideal obligation of \cref{sec:precondition:problem}.

\textbf{Discharging preconditions}
When proving refinements involving modules with precondtions, we ensure that none of them enter into \NONE{} via enforcing their internal states stay as \SOME{} in the simulation invariant.

For our top-level theorem (\cref{thm:main}), $\RISCVPROM{}$ is a regular module without precondition wrappers, hence whether it can go into \NONE{} is moot.

\end{document}